\documentstyle[aps,epsfig,multicol,fancyheadings]{revtex}
\begin{document}

\title{Roughness exponent in the fracture of fibrous
materials}

\author{I.\ L.\ Menezes-Sobrinho}

\address{Departamento de F\'{\i}sica, Universidade Federal de Vi\c cosa,
36570-000 Vi\c cosa, MG, Brazil} \date{\today}

\maketitle

\begin{abstract}
In this paper, a computational model in (2+1)-dimensions which simulates the 
rupture process of a fibrous material submitted to a constant force $F$, is
analyzed. The roughness exponent $\zeta$ at the boundary that separates two
failure regimes, catastrophic and slowly shredding, is evaluated. In the
catastrophic (dynamic) regime the initial strain creates a crack which
percolates rapidly through the material. In the slowly shredding (quasi-static)
regime several cracks of small size appear in all parts of the material, the
rupture process is slow and any single crack percolates the sample. At the
boundary between these two regimes, we obtained a value $\zeta\simeq 0.42\pm
0.02$ for the roughness exponent, in agreement with results provided by other
simulations in three dimension. Also, at this boundary we observed a power law
behavior on the number of  cracks versus its size.  \end{abstract}

\noindent {\it PACS \# \ \ 62.20.Mk, 64.60.Fe, 05.40.+j}

\begin{multicols}{2}[]
\narrowtext
\section{Introduction}\label{intro}
The fracture process in disordered materials is a subject of 
intensive research and has attracted much scientific and industrial
interest \cite {livro,zape,maes}. The fracture process is extremely
sensitive to disorder, which may act as obstacles difficulting  the
propagation of cracks through the material. Thus, disorder has a strong
influence on the roughness of the fracture surface. Several computational
models have been constructed to study the phenomenon of fracture of these
materials, such as, the extensively studied  fuse \cite{arc} and the
well-known fiber bundle models \cite{zhang,dux,isma1,isma,ze}, created from
the pioneer work of Daniels \cite{dan}.

Experiments have shown that the fracture surface in disordered materials
can often be described by self-affine scaling \cite{mand,dag,lop}. In this
case, the roughness of the fracture profile can be characterized by a roughness
exponent $\zeta$. Some experimental works have claimed that the roughness
exponent $\zeta$ has a universal value of 0.8 \cite{bou,maloy,sch}. However
this universality was questioned by Milman {\it et al}. \cite{mil}, which
experimentally found a roughness exponent closer to 0.5.
From the theoretical point of view, numerical models have been searched
in order to evaluate the roughness exponent $\zeta$. Simulations have 
shown that $\zeta\sim 0.7$ in two dimensions  \cite{hansen,g} and that
$\zeta$ ranges from $0.4$ to $0.5$ in three dimensions \cite{rai,bat}.
Nowadays there is a conjecture relating the smaller and the higher value of the
roughness exponent $\zeta$ to the speed of crack propagation
through the sample \cite{bat}. The greater value has been associated with a
high speed of crack propagation and interpreted as a dynamic regime. In
contrast, the smaller value of the exponent was related with a quasi-static
regime, where the dynamic effects of the propagation are negligible.
Experimentally, the smaller and the higher value of $\zeta$ are connected to
the length scale at which the crack is examined. The smaller value is
associated to small length scales and the higher value is connected to
large length scales \cite {bou,bou1}. Several models used to study the
characterization of the fracture surface by the roughness exponent do not
consider the influence of the temperature on the fracture process. In
this paper we present a fibrous model in (2+1) dimensions for which it is
possible to obtain the fracture profile of a fibrous material submitted to a
constant force F, for example, by a hanging weight on it. In this work the 
influence of the temperature $t$ on the fracture process was considered. We
show that, at the boundary between the catastrophic and the slowly shredding
regime, the roughness exponent $\zeta$ does not depend on the temperature $t$. 

\section{Model}\label{num}
Our model consists of a bundle of
$N_0=L\times L$ parallel fibers, all with the same elastic constant,
$k$, distributed on a triangular lattice. In order to simulate the height of
the sample, the fibers are divided in $\eta$ segments with the same length. The
fiber bundle is fixed at both extremes by two parallel plates, one of these is
fixed and in the other a constant force $F$ is applied. This force is equally
and completely distributed in the fiber bundle, submitting all fibers to the
same linear deformation $z=F/Nk$, where $N$ is the number of unbroken
fibers. This type of distribution is called equal load sharing (ELS)
\cite{zhang,dux,kloster}. Another type is the local load sharing
(LLS), where the load of a broke fiber is transfer onto its two nearest
unbroken fibers \cite{newman,dong}.  
At our model, when the deformation $z$  reaches a critical value $z_c$, the
failure probability of an isolated fiber is equal to one. The failure
probability  of a fiber {\it i} is given by \cite{isma}  \begin{equation}
P_i(\delta,t)={\delta\over(n_i+1)}\exp\left[{(\delta^2-1)\over t}\right],
\label{eq1} \end{equation}
where $n_i$ is the number of unbroken neighboring fibers,
$\delta=z/z_c=F/Nkz_c$ is  the strain of the material, $t=K_BT/E_c$ is the
normalized temperature, $K_B$ is the Boltzmann constant, $T$ is the
absolute temperature and $E_c$ is the critical elastic energy. In this model,
besides finding the failure probability of a fiber, we have to indicate in
which segment it breaks. Since, each fiber of the bundle can break at
different parts a fracture surface is produce. Similar procedure was used in
Ref \cite{pre} to obtain the fracture profile of a model for fracture on
fibrous materials in (1+1) dimensions.  The segment is randomly selected and
the probability of the fiber to break in it is given by \cite{pre}. 
\begin{equation} \phi_j(m_j)={(m_j+1)\over g}, \label{eq2} \end{equation}
where $m_j$ is a vector which indicates how many times a segment $j$
broke and $g=\sum_j (m_j+1)$. Eq.~(\ref{eq2}) simulates 
a concentration of tension near to the region where the fiber bundle is
weaker.

At the beginning of the simulation, the bundle is submitted to an initial strain
given by 
\begin{equation}
\delta_0={z_o\over z_c}={F\over {N_0kz_c}}.
\label{eq3}
\end{equation}
 At each time step we randomly choose a fiber of a set of $N_q =qN_o$ unbroken
fibers. The number $q$ represents a percentage of fibers and allow
us to work with any system size. Then, using Eq.~(\ref{eq1}), we evaluate
the fiber failure probability $P_i$ and compare it with a random
number $r$ in the interval [0,1). If  $r<P_i$ the fiber breaks. We then choose
a segment $j$ in the fiber and evaluate its probability $\phi$ to break, 
using Eq.~(\ref{eq2}). If the  probability $\phi$ is
higher than a random number $f$ the fiber breaks in the chosen segment.
If not, we analyze the neighboring segments (j+1) and (j-1) and again,
evaluate the probability $\phi$. If the condition $f<\phi$ does not hold to
neither of the  neighboring segments, we return to the initial segment and test
the condition  $f<\phi$ for a new value of $f$. This process continues
until the  condition $f<\phi$ is true. Once defined the segment where the
fiber breaks, we begin to test all neighboring unbroken fibers. 
The first
segment tested in the neighboring unbroken fibers is the one in which the
previous fiber broke. The failure probability $P_i$ of these neighboring
fibers increases due to the decreasing of $n_i$ and a cascade of breaking
fibers may begin. This procedure describes the propagation of a crack through
the fiber bundle in the perpendicular direction to the applied force. 
The process of propagation stops when the
test of the probability does not allow rupture of any other fiber on the
border of the crack or when the crack meets another already formed crack. The
same cascade propagation is attempted by choosing another fiber of the set
$N_q$. After all the $N_q$ fibers have been tested, the strain $\delta$ is
increased if some fibers have been broken. Since the force is fixed, the
greater the number of broken fibers, the larger is the strain on the intact
fibers and the higher is their failure probability. Then, another set of $N_q$
unbroken fibers is chosen and all the rupture process is restarted. The
simulation terminates when all the fibers of the bundle are broken, {\it
i.e.}, when the bundle is divided into two parts.

\section{Results}\label{scale}
We performed simulations considering $L=2000$ ($N_0=4\times 10^6$ fibers), the
elastic constant $k=1$, the critical deformation $z_c=1$ and the number of
segments $\eta=1000$.

The failure probability (Eq. \ref{eq1}) can be written as

\begin{equation}
\label{fp1}
P_i(z)= {\Gamma (t,\delta) \over {(n_i +1)}  },
\end{equation}

where the parameter $\Gamma (t,\delta)$ is defined as

\begin{equation}
\label{po}
\Gamma (t,\delta) = \delta\exp \left({\delta ^2 - 1} \over t \right).
\end{equation}

For a triangular lattice (with coordination number 6) and 
$\Gamma (t,\delta) \geq 6$, the rupture of any fiber induces
the rupture of the whole bundle, i.e., the bundle 
 breaks with just one crack. Obviously, this crack forms a cluster
 which percolates through 
the entire system.

We can define the density of the percolation crack as

\begin{equation}
\label{rho}
\rho = {N_{pc}\over N_0},
\end{equation}
where $N_{pc}$ is the number of broken fibers belonging
to the percolating crack. Thus, when $\Gamma (t,\delta) \geq 6$, we
have $\rho = 1$.

\begin{figure}[f]
\centerline{\epsfig{file=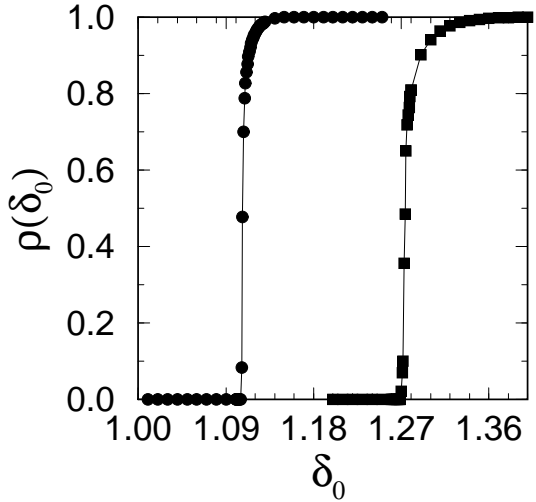,width=7cm,angle=-0}}
\caption{Density of the percolating cluster $\rho$ vs the initial
strain $\delta_0$ for two different temperatures: $t=0.5$ (circles) and
$t=2.0$ (squares). The data were averaged over 1000 statistically
independent samples.}   
\label{is1}   
\end{figure}

Figure \ref{is1} shows the density of the percolation crack $\rho$ versus
the  initial strain $\delta_0$, for two different 
temperatures. Notice that, for high values of $\delta_0$, $\rho=1$ and for low
values of $\delta_0$ the density of the percolating cluster $\rho$ jumps to
zero. Thus, we may assume that there is a critical value $\delta_{0c}$ which
depends on the temperature $t$. For the temperatures $t=0.5$ and $t=2.0$ used
in our simulation the values obtained for $\delta_{0c}$ are $1.11$ and $1.27$
respectively. Above $\delta_{0c}$ there is a percolation crack and below it any
single crack percolates the fiber bundle. The critical value, $\delta_{0c}$,
represents the transition between two failure regimes \cite{isma1}:
catastrophic and slowly shredding. In the catastrophic regime there are cracks
that percolate the fiber bundle and in the slowly shredding any crack
percolates the bundle. In Ref.\cite{isma1} we have shown that these two regimes
are separated by a second order phase transition and determined from (Eq.
\ref{po}) the critical line separating these two failure regimes in the plane
$t\times \delta_0$.

\begin{figure}[f]
\centerline{\epsfig{file=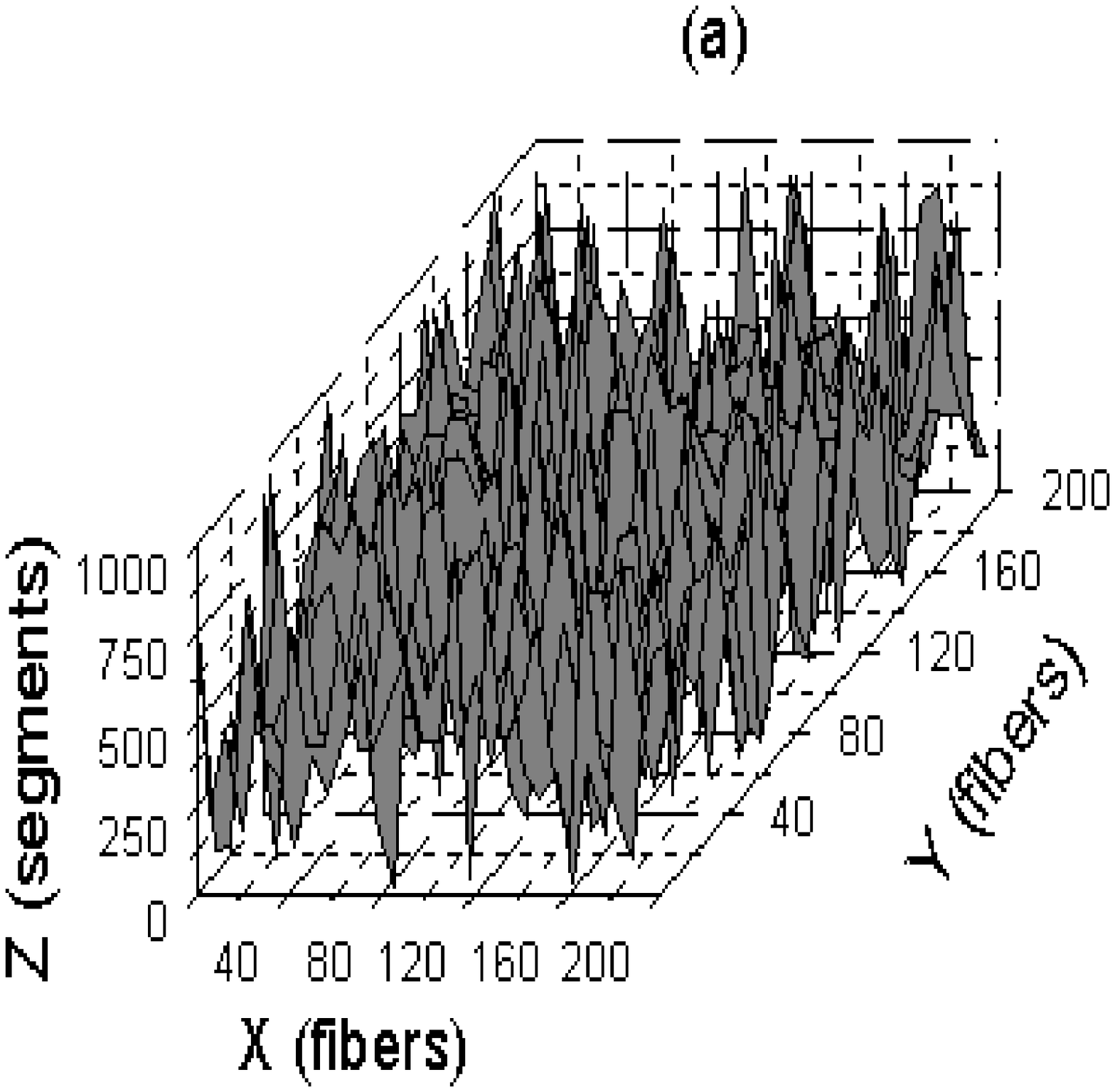,width=7cm,angle=-0}}
\centerline{\epsfig{file=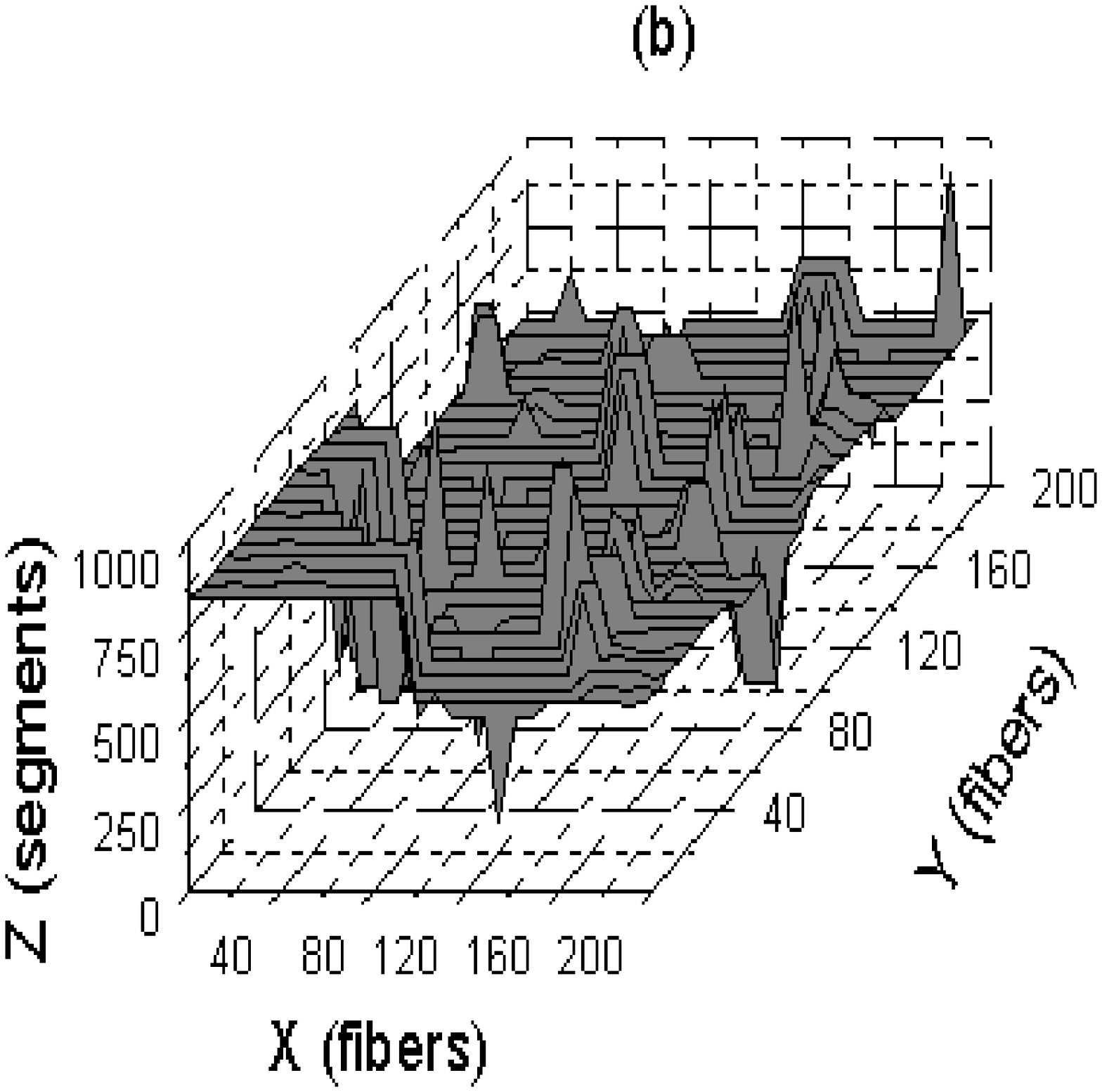,width=7cm,angle=-0}}
\centerline{\epsfig{file=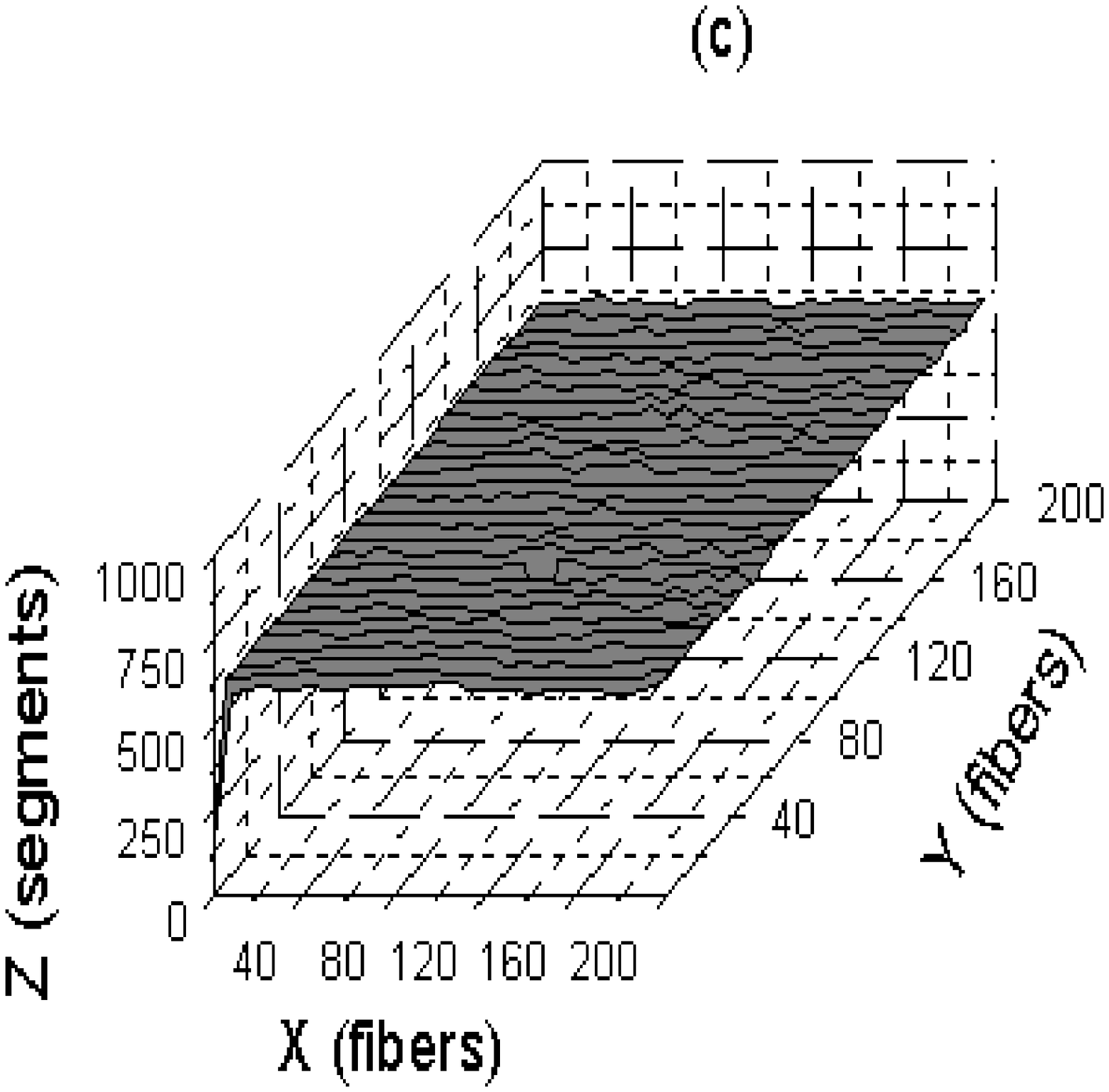,width=7cm,angle=-0}}
\caption{Fracture surface for three different initial strains. In (a) 
we have: $\delta_0=0.4$, in (b) $\delta_0=1.27$ and in (c) $\delta_0=1.4$. In
this particularly simulation we have used a total of $N_0=4\times 10^4$
fibers.}   \label{is2}  
\end{figure}

Figure \ref{is2} shows the fracture surface obtained for $t=2.0$ and three
different initial strains $\delta_0$. In Fig. \ref{is2} (a) the fracture
surface is very rough and this profile is characteristic of a shredding
fracture, in which the speed of crack propagation is low due to a slow process of
successive rupture of fibers in the material. In Fig. \ref{is2} (c) the
fracture surface presents little roughness and is characteristic of
catastrophic fracture. Here the crack propagates with high speed and the
breakage of a single fiber induces the rupture of the whole the bundle. For
$\delta_0=1.27$ [Fig. \ref{is2} (b)] the fracture occurs at the boundary
between the two failure regimes. It can be seen from Fig. \ref{is2} that the
rupture of the sample begins at different segments, since we did not used a
deterministic starting notch in our simulations. 

In order to evaluate the roughness exponent $\zeta$ different
one-dimensional cuts in the fracture surface were considered. The roughness $W$
of each cut was found by the method of the best linear least-square fitting
described in \cite{jaf}. In this method, the roughness $W(\epsilon)$ in the
scale $\epsilon $ is given by 
\begin{equation}
 W(\epsilon)={1\over M}\sum_{i=1}^{M}w_i(\epsilon)
\end{equation}
and the local roughness $w(\epsilon)$ is defined as
\begin{equation}
w^2={1\over(2\epsilon +1)}\sum_{j=i-\epsilon}^{i+\epsilon}
[h_j-(a_ix_j+b_i(\epsilon))]^2.
\end{equation}
$a_i(\epsilon)$ and $b_i(\epsilon)$ are the linear fitting coefficients to the
displacement data on the interval $[i-\epsilon,i+\epsilon]$ centered on the
fiber $i$.

\begin{figure}[f]
\centerline{\epsfig{file=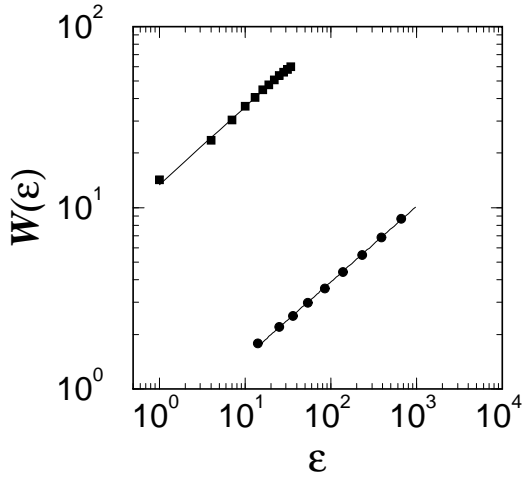,width=7cm,angle=-0}}
\caption{Log-Log plot of the roughness $W$ as function of the scale $\epsilon$
for two different temperatures: $t=0.5$ (circles) and $t=2.0$ (squares). The
two straight lines have slope 0.42.}  
\label{is3} 
\end{figure}

\begin{figure}[f]
\centerline{\epsfig{file=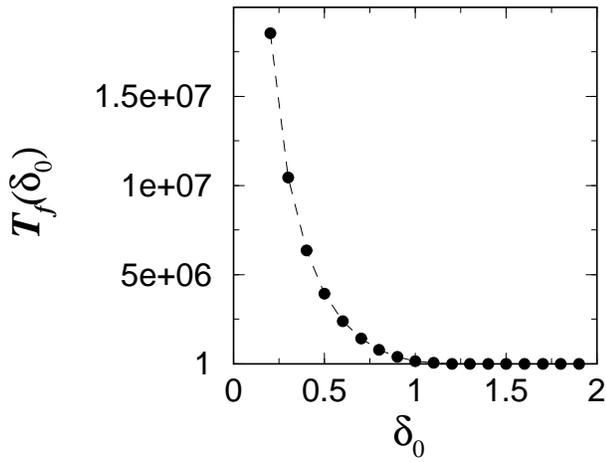,width=8cm,angle=-0}}
\caption{Plot of the time to failure (in Monte Carlo step) $T_f$ vs the initial
strain $\delta_0$ for $t=2.0$. The data were averaged over 1000 statistically
independent samples.}   
\label{is5}   \end{figure}

The roughness exponent $\zeta$ at the critical point $\delta_{0c}$  for
two temperatures $t$ was calculated and a value of $\zeta\simeq
0.42\pm 0.02$ was obtained. Our results indicate that at the critical point the
value of $\zeta$ does not depend on the temperature $t$. Figure \ref{is3}
shows the fits for roughness $W$ at $t=0.5$ and $t=2.0$. We also verified 
that, as the
initial strain $\delta_0$ decreases below $\delta_{0c}$, the rupture process
and, consequently, the speed of crack propagation become more
slow. In this situation the roughness exponent $\zeta$ tend to zero. 
In Fig. \ref{is5} we show the plot of the
time to failure (in Monte Carlo step) $T_f$ as function of the initial strain
$\delta_0$ for $t=2.0$. Notice that the time decreases with increase of the
initial strain $\delta_0$. For $\delta_0>\delta_{0c}$ the rupture is
catastrophic and the material breaks in the first time step. In this regime was
not possible to find the roughness exponent $\zeta$. We believe that in this
regime the fracture surface ceases to be self-affine. 

The roughness exponent $\zeta$ was calculated along the
\^{x} and \^{y} direction. At both directions we verify that the roughness
exponent has the same value. This behavior is expected for a large variety of
materials, where the two directions have similar scaling properties \cite
{plo,parisi}. At the transition we have observed a power law behavior on the
number of  cracks versus its size. 

\begin{figure}[f]
\centerline{\epsfig{file=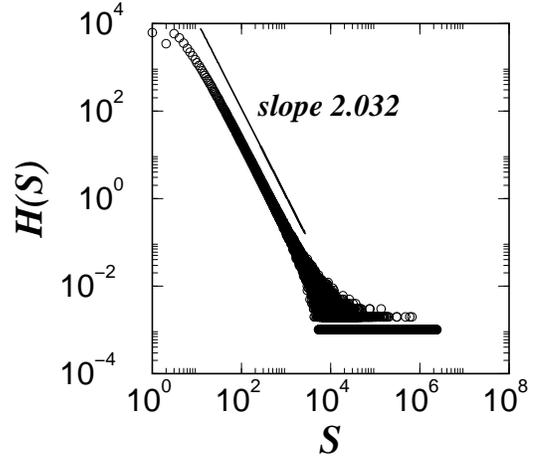,width=7cm,angle=-0}}
\caption{Log-Log plot of the frequency of the cracks $H_s$ vs the
size of the cracks $S_c$ for $t=2.0$ and $\delta_{0c}=1.27$. The straight line
has a slope $2.032$. The data were averaged over 1000 statistically independent
samples.}    \label{is4}   \end{figure}

The log-log diagram of the frequency of 
the cracks $H_c$ versus their sizes $S_c$ observed for $t=2.0$ and
$\delta_{0c}=1.27$ is shown in Fig. \ref{is4}. Note that at the beginning
$H_c$ seems to decay linearly so that we can assume a  power law 
\begin{equation} H_c\sim S^{-\alpha}, \end{equation}
where $\alpha=2.032\pm 0.007$. We observed that at the transition this
exponent does not depend on the temperature. 

\section{Conclusion}\label{discussion}
In conclusion, we have studied a model for fracture on fibrous materials in
(2+1) dimensions which provides the fracture surface of the material,
in contrast with previous models. We calculated the roughness exponent $\zeta$
and showed that it does not depend on the temperature $t$. Our results
indicate a value of $\zeta\simeq 0.42\pm0.02$ at the boundary between the
catastrophic (similar to the dynamic regime) and the slowly shredding regime
(similar to the quasi-static regime). This value is the same obtained in other
type of simulations in three dimensions \cite{rai,bat} that related the
exponent $\zeta$ with the quasi-static regime. We also have shown that at the
boundary there is a power law connecting the frequency $H_c$ and
the size $S_c$ of the size. This power law behavior is characteristic of
systems in the criticality and is a good indication that the fracture surface
is a fractal.

It is a pleasure to acknowledge Professors A. T. Bernardes and J. G. Moreira
for precious guidance and many interesting discussions. We also thank M.L
Martins and M. S. Couto for hel\-pful cri\-ti\-cism of the manuscript and the
kind hospitality of the Departamento de F\'{\i}sica, UFMG. The author also
acknowledges the FAPEMIG (Brazilian agency) for financial support.   


\end{multicols}
\widetext
\end{document}